\relax
\documentclass[letterpaper]{article} 
\usepackage{aaai19}  
\usepackage{times}  
\usepackage{helvet}  
\usepackage{courier}  
\usepackage{url}  
\usepackage{graphicx}  
\usepackage{array,multirow}
\usepackage{xcolor}
\usepackage{array}
\usepackage{booktabs}
\usepackage{enumitem,kantlipsum}
\usepackage[normalem]{ulem}
\usepackage{soul}
\usepackage{natbib}
\usepackage{xspace} 

\renewcommand{\prec}{\ensuremath{\mathcal{P}}\xspace}
\newcommand{\recall}{\ensuremath{\mathcal{R}}\xspace}
\newcommand{\fscore}{\ensuremath{\mathcal{F}}\xspace}
\newcommand{\accuracy}{\ensuremath{\mathcal{A}}\xspace}

\newcolumntype{L}[1]{>{\raggedright\let\newline\\\arraybackslash\hspace{0pt}}m{#1}}
\newcolumntype{C}[1]{>{\centering\let\newline\\\arraybackslash\hspace{0pt}}m{#1}}
\newcolumntype{R}[1]{>{\raggedleft\let\newline\\\arraybackslash\hspace{0pt}}m{#1}}
\begin{document}
%
\title{A Few Topical Tweets are Enough for Effective Stance Detection}
\author{
Younes Samih and Kareem Darwish\\
Qatar Computing Research Institute\\
Hamad bin Khalifa University, Doha, Qatar\\
\{ysamih,kdarwish\}@hbku.edu.qa
}

\maketitle
\begin{abstract}
Stance detection entails ascertaining the position of a user towards a target, such as an entity, topic, or claim. Recent work that employs unsupervised classification has shown that performing stance detection on vocal Twitter users, who have many tweets on a target, can yield very high accuracy (+98\%). However, such methods perform poorly or fail completely for less vocal users, who may have authored only a few tweets about a target.  In this paper, we tackle stance detection for such users using two approaches.  In the first approach, we improve user-level stance detection by representing tweets using contextualized embeddings, which capture latent meanings of words in context. We show that this approach outperforms two strong baselines and achieves 89.6\% accuracy and 91.3\% macro F-measure on eight controversial topics. In the second approach, we expand the tweets of a given user using their Twitter timeline tweets, and then we perform unsupervised classification of the user, which entails clustering a user with other users in the training set. This approach achieves 95.6\% accuracy and 93.1\% macro F-measure. 

\end{abstract}

\section{Introduction}

\begin{figure}
    \centering
    \includegraphics[width=.7\linewidth]{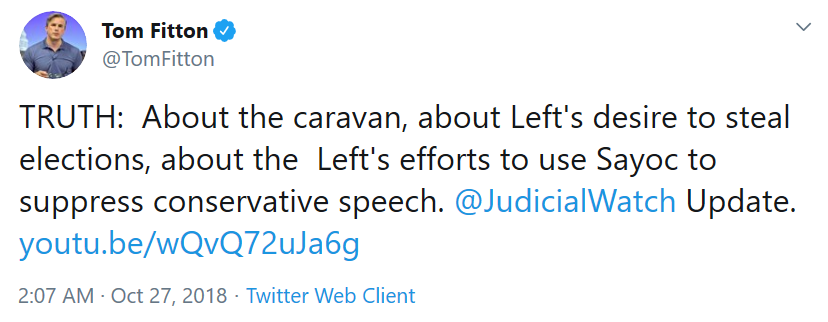}
    (a)
    \includegraphics[width=.7\linewidth]{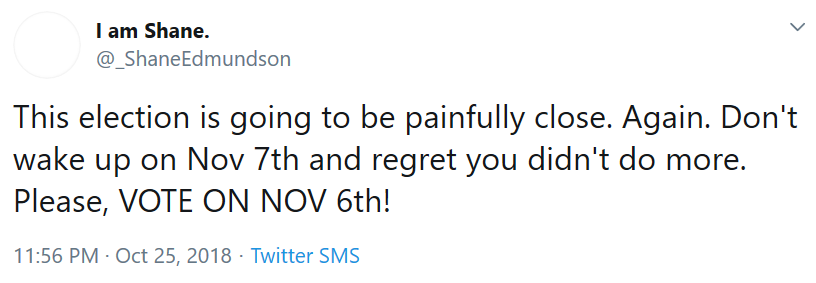}
    (b)
    \caption{Sample 2018 US midterm election related tweets that either express a very clear stance (a) or not (b)}
    \label{fig:sampleTweets}
\end{figure}

Stance detection entails identifying the position of a user towards a topic, an entity, or a claim \citep{mohammad2016semeval}.  Effective stance detection, particularly in the realm of social media, can be instrumental in gauging public opinion, identifying intersecting and diverging groups, and understanding issues of interest to different user communities \citep{magdy2016isisisnotislam}. Much recent works have explored varying stance detection methods including supervised, semi-supervised, and unsupervised user classification \citep{darwish2019unsupervised,magdy2016isisisnotislam,pennacchiotti2011democrats,wong2013quantifying}, and much of the work has focused on stance detection for Twitter users.  The different approaches have advantages and disadvantages.  For example, supervised methods are simple to implement, but they require manually annotated training data and their accuracy varies widely based on classification features, the classification technique, and the number of training and test examples \citep{magdy2016isisisnotislam}.  Though semi-supervised and unsupervised methods typically use user interactions and often may yield perfect classification, they are effective in classifying highly vocal users with many topical tweets \citep{darwish2019unsupervised}. Most of these methods produce sub-optimal results for users who rarely express their opinion, and for whom we may only have one or two topically related tweets.  Though a single tweet might be explicitly clear, often it may lack sufficient context to determine the stance of the user. Figure \ref{fig:sampleTweets} show two tweets that pertain to the 2018 US midterm elections, where the first expresses a lucid pro-republican stance and the second could have been authored by a supporter of either the republican or democratic party.  In this paper, we aim to effectively identify the stance of Twitter users towards specific targets (entities or topics), where the users have mentioned the targets in only a few tweets (less than two tweets on average). 

To do so, we employ two approaches.  In the first approach, we classify user based on their tweets that are represented using contextualized embeddings, which capture latent meanings of words in context. Specifically, we use BERT embeddings to represent tweets, and we fine tune the embeddings for every topic. We compare this approach to two strong baselines, namely using Support Vector Machine (SVM) classification and fastText, which is a deep learning based classifier. In the second approach, we expand the tweets of a given user using their Twitter timeline tweets, and then we perform unsupervised classification of the user by clustering him/her with the users in the training set. Using such expansion allows us to make use of echo chambers that form on Twitter, where users with similar views tend to retweet similar accounts beyond the topic at hand.  To test our approaches, we used a dataset containing tweets on 8 polarizing US-centric topics. We also examine the effect of expansion when using SVM, fastText, and contextualized embeddings. For testing, we randomly selected 100 users for each topic that have less than 5 topical tweets, and we manually labeled them for stance.  To construct the training set, we used unsupervised stance detection to automatically label the most active 5,000 users per topic, and for every topic we used a balanced set of 500 users per stance as our training set \citep{darwish2019unsupervised}.  Since the approaches rely on different features and utilize different classification techniques, we indicate which approach works best under different conditions. 

The contributions of this paper are as follows:
\begin{itemize}
    \item We fine-tune contextualized embeddings to generate latent representations of tweets to effectively classify the stance of users based on only one or two tweets.  We achieve an of 89.6\% and F-measure of 91.3\%, which are significantly higher than the scores we achieved for both baselines.
    \item We show that using additional timeline tweets for the users that we wish to classify, and then using unsupervised classification, where we cluster the test user with users in our training, leads to an accuracy of 95.6\% and F1-measure of 92.0\%.  In doing so, we extend prior work on unsupervised stance detection to effectively classify both users who are vocal on a topic as well as those with perhaps one or two topical tweets. 
    \item We conduct error analysis on our best setups to determine the sources of the errors, which would help in guiding the choice of classification method.
    \item We plan to release the tweet IDs of the test set along with the associate gold labels. Further, we plan to release the code that performs classification based on contextualized embeddings. 
\end{itemize}

\section{Related Work}
Over the last few years, much research has focused on stance detection. The goal of stance detection is to ascertain the positions of users towards some target such as a topic, person, or claim ~\citep{thomas:2006,mohammad:2016,barbera2015birds,barbera2014understanding,borge2015content,cohen2013classifying,colleoni2014echo,conover2011predicting,fowler2011causality,himelboim2013birds,magdy2016isisisnotislam,magdy2016failedrevolutions,makazhanov2014predicting,weber2013secular}. \\ While stance may easily be detected by humans, machine learning models often fall short, particularly for users who talk about a target sparingly. Several studies have focused on modeling stance by introducing different features ranging from linguistic and structural features~\citep{mohammad:2016} all the way to network interactions and profile 
information \citep{borge2015content,magdy2016isisisnotislam,magdy2016failedrevolutions,weber2013secular}. Much work on stance detection involved using supervised and semi-supervised classification methods. One of the major downsides of both classification methods is the need for a seed list of manually labeled users, which is time consuming and requires topic expertise.  Supervised learning is sensitive to the classification features, the size of the training sets, the number of available tweets for users in the test set, and the classification algorithm \citep{borge2015content}. Some common classification features include: lexical, syntactic, and semantics feature; network features such as retweeted accounts and user mentions; content features such as words and hashtags; and user profile information such as name and location \citep{Aldayel2019,magdy2016isisisnotislam,magdy2016failedrevolutions,pennacchiotti2011democrats}. 
Some commonly used classification algorithms include SVMs and deep learning classification~\citep{zarrella2016mitre}. \cite{stancy2019} present a neural network model for stance classification by augmenting BERT representations with a novel consistency constraint to determine stance with respect to both a claim and perspective.  We extend their work in two ways, namely: we drop the need to have a claim and perspective, and we couple BERT supervised classification with unsupervised classification to effectively tag vocal and non-vocal users. 
Semi-supervised methods such as \textit{label propagation} \citep{barbera2015birds,borge2015content,weber2013secular} often rely on two users retweeting identical accounts or tweets to propagate a label of one user to another.  Though such typically achieves high precision (often above 95\%) \citep{darwish2018predicting}, it is generally successful in tagging vocal users with strong opinions. Recently, \cite{darwish2019unsupervised} have introduced a highly effective unsupervised method for predicting the stance of prolific Twitter users towards controversial topics. By projecting users onto a low-dimensional space and then clustering them allows for clear separation between vocal users with respect to their stance~\citep{darwish2019unsupervised}. This method confers two main advantages over previous methods, namely: it does not require any initial manual labeling, and classification accuracy is nearly perfect.  However, it is successful in labeling vocal users only and fails on users with very few topical tweets.  
We extend prior work on  unsupervised  stance  detection  to  effectively  classify both prolific and non-prolific users in a holistic way by aggregating both supervised and unsupervised methods. Further, we extend upon prior deep-learning based supervised classification to use contextual embeddings that capture syntactic and semantic features of words in context.

\section{Data Sets}
\subsection{Topics}
Our dataset includes tweets on eight polarizing topics that are US-centric, which were graciously provided to us by Stefanov et al. \citep{stefanov2019predicting}.  Table \ref{tab:topics} lists all the topics including when the tweets were collected and the number of tweets per topic.  The topics include both long-standing issues such as gun control and transient issues such as the nomination of Judge Kavanaugh to the US Supreme Court.  There is also a non-political issue, namely vaccination.  The tweets were also filtered based on user-stated locations to limit our data to US users. The filtering was done using a gazetteer that includes either US (or its variants) and state names (and their abbreviations).

\begin{table}[ht]
    \centering
    \scriptsize
    \begin{tabular}{p{3.7cm}|p{2.3cm}|p{.8cm}}
    Topic & Date Range & Tweets \\ \hline
    Climate change & Feb 25--Mar 4, 2019 & 1,284,902 \iffalse6,535,863\fi \\ \hline
    Gun control &  Feb 25--Mar 3, 2019 & 1,782,384 \iffalse7,583,311\fi \\ \hline
    Ilhan Omar (remarks on Israel lobby) & Mar 1--9, 2019 & 2,556,871 \\ \hline
    Immigration & 2,341,316 \iffalse7,593,058\fi \\ \hline
    Midterm (elections 2018) & Feb 25--Mar 3, 2019 & 2,564,784 \iffalse8,369,133\fi \\ \hline
    Kavanaugh (nomination to Supreme Court) & Sept. 28-30 \& Oct. 6-9, 2018 & 2,322,141 \\ \hline
    Vaccination & Mar 1--9, 2019 & 301,209 \\ \hline
    \end{tabular}
    \caption{\label{tab:topics}Controversial topics used in study.}
\end{table}

\subsection{Training Set}

Given the tweets for every topic, we performed per topic unsupervised stance detection \citep{darwish2019unsupervised}.  This approach identifies the most active $n$ users per topic and computes similarity between them based on a common feature, such as which hashtags they use or which accounts that they retweet.  Next, the users are projected onto a lower dimensional space in a manner where similar users are brought closer together and dissimilar users are pushed further apart. Then the projected users are clustered.  Using the best reported setup \citep{darwish2019unsupervised}, we used the most active 5,000 users with at least 10 tweets, computed similarity between them based on which accounts they retweeted, projected users using UMAP \citep{mcinnes2018umap}, and clustered them using the mean shift clustering algorithm. Stefanov et al. \citep{stefanov2019predicting} estimated that the accuracy of their approach on these topics to be 98\%. Next we took 500 random users from the two largest clusters to construct a balanced training set, and we manually inspected a few users from each cluster to give an overall label to each cluster (ex. pro- or anti- gun control).  Further, we crawled the timeline tweets of the users in our training set.

\subsection{Test Set}
For each topic, we randomly selected 200 users who have less than 5 tweets.  The average number of tweets per user ranged across topics between 1.25 and 1.77 tweets.  An annotator who is well versed with US politics manually examined the per topic tweets of users to determine their stances.  If the tweets of a user were not sufficient to ascertain their stance, the annotator manually searched and examined their tweets on Twitter in an effort to find further clues.  If no conclusive evidence of stance were found, the annotator skipped the user.  The annotator labeled up to 100 users per topic.  Next, we scraped the timeline of all the labeled users.  Due to the time difference between collecting topical tweets and when we initiated the scraping of users' timelines, some user accounts were either deleted, suspended, or made protected.  Table \ref{tab:labeledUsers} lists the number of labeled users and the subset of them for whom we were able to scrape their timelines. Since there is a disparity between the number of manually labeled users with and without timeline tweets, we shall report separately on them in all experiments. Specifically, we put users for whom we were not able to collect timeline tweets into Set A, and we put the remaining users in Set B.

\begin{figure}
    \centering
    \includegraphics[width=.9\linewidth]{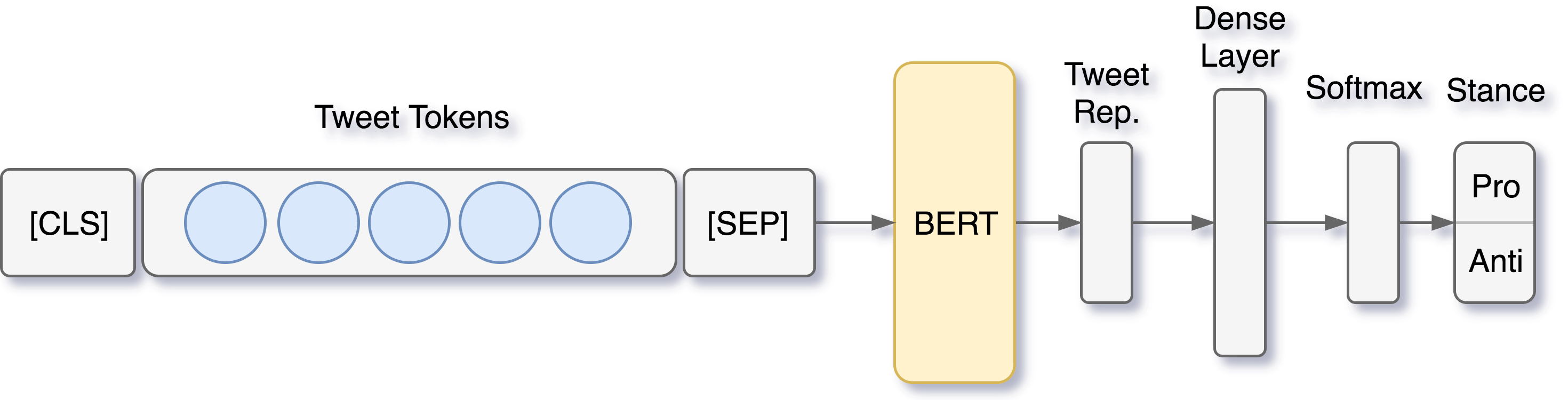}
    \caption{Fine-tuning BERT for stance classification.}
    \label{fig:bert}
\end{figure}

\begin{table}[]
    \centering
    \scriptsize
    \begin{tabular}{l|c|c}
         Topic & Labeled Users & User Timelines Scraped \\ \hline
         Climate change & 100 & 54\\
         Gun control & 58 & 26 \\
         Ilhan Omar & 100 & 39\\
         Immigration & 100 & 43\\
         Midterm & 100 & 45\\
         Racism \& police & 100 & 54\\
         Kavanaugh & 100 & 55\\
         Vaccines & 100 & 57\\
    \end{tabular}
    \caption{Per topic labeled users in test set along with the number of users for which were able to scrape their timelines}
    \label{tab:labeledUsers}
\end{table}

\section{Classification Models}
\subsection{Supervised Classification}
\subsubsection{Baselines}
As baselines, we used two different classification methods, namely Support Vector Machines (SVM) classifier and a deep learning based text classifier.  For the SVM classifier, we used the SVM$^{Light}$ implementation with a linear kernel with default parameters \citep{Joachims02svm}\footnote{\url{http://svmlight.joachims.org/}}.  We employed two feature types, namely: the accounts that users retweeted; and the words in tweets, including retweeted accounts, hashtags, and user mentions and replies. Prior work has shown that using retweeted accounts as features yields better results compared to using the content of tweet \citep{darwish2018predicting}.  When using words in tweets, we tokenized tweets using NLTK, removed all URLs and emoticons, retained all hashtags and user mentions, and specifically delineated retweeted accounts by adding 'RT\_' before them. We chose to distinguish between retweeted accounts and user mentions because retweeting commonly signifies agreement and user mentions (including replies) may indicate opposition.  We concatenated the aforementioned features from all the tweets of a user, and we constructed a feature vector, where the value of each unique feature was set to its frequency across all tweets of a user.  For the deep learning based classifier, we used fastText, which is an efficient text classifier that has been shown to be effective for different text classification tasks \citep{joulin2016bag}.  Since fastText was designed for sentence-level classification, we opted to perform tweet-level classification. During training, we assigned the label of a user to all his/her tweets. During testing, we averaged per class confidence scores across all tweets for a user, and we assigned the label with the highest average confidence to the user. As for features, we used all the words in tweets, and we preprocessed tweets in the manner described earlier for SVM.  We opted not use retweeted accounts only as the number of retweeted accounts was arbitrary for each user and fastText is not well suited for long input text.

\paragraph{Contextualized Embeddings}
Over the last several years, pre-trained embedding \citep{mikolov2013,pennington2014} have helped achieve significant improvements in a wide range of classification tasks in natural language processing. Representing words as vectors in a low-dimensional continuous space and then using them for downstream tasks lowered the need for extensive manual feature engineering. However, these pre-trained vectors are static and fail to handle polysemous words, where different instances of a word have to share the same representation regardless of context. More recently, different deep neural language models have been introduced to create contextualized word representations that can cope with the issue of polysemy and the context-dependent nature of words. Models such as OpenAi GPT~\citep{radford2018}, ELMo~\citep{Peters2018}, BERT (Bidirectional Encoder Representations from Transformers)~\citep{devlin-2019-bert}, and UMLFIT~\citep{howard-ruder-2018}, to name a few, have achieved ground-breaking results in many NLP classification and language understanding tasks. For this paper, we use BERT\textsubscript{base-multilingual}\footnote{We also experimented with different contextualized embedding, such as RoBERTa~\citep{roberta2019}, Albert~\citep{albert2019}, and XLM~\citep{xlm2019} and BERT\textsubscript{base-multilingual} performed the best. } (referred to hereafter simply as BERT), which we fine-tune for stance detection, as this eliminates the need for heavily engineered task-specific architectures. BERT is pre-trained on Wikipedia text from 104 languages and comes with hundreds of millions of parameters. It contains an encoder with $12$ Transformer blocks, hidden size of $768$,  and $12$ self-attention heads. As shown in Fig. \ref{fig:bert}, We fine-tuned BERT by adding a fully-connected dense layer followed by a softmax output layer, minimizing the binary cross-entropy loss function for the training data. For all experiments, we used HuggingFace\footnote{\url{https://github.com/huggingface/transformers}} transformer implementation with PyTorch\footnote{\url{https://pytorch.org/}} as it provides pre-trained weights and vocabulary.  As for features, we used all the words in tweets that were preprocessed in the manner described earlier for SVM and fastText.  Similar to fastText, we performed tweet-level classification, and we used the average softmax output scores per class across all tweets for a user to assign a label to a test user.

\paragraph{Unsupervised Classification}
For unsupervised classification, we used the same unsupervised classification method described earlier, which we used to prepare the training set.  Specifically, we constructed a feature vector for each test user based on the accounts he/she retweeted, computed its similarity to all users in the training set, projected all the users in the training along with the test user into a lower dimensional space using UMAP, and lastly clustered the users using mean shift.  We then labeled the test user using the majority label of the cluster in which the user appeared.

\begin{table*}[ht]
\scriptsize
    \centering
    \begin{tabular}{l|c|c|c|c||c|c|c|c||c|c|c|c||c|c|c|c} 
    \multicolumn{17}{c}{Set A} \\ 
	&	\multicolumn{4}{c}{fastText}		&	\multicolumn{4}{|c}{SVM$_{RT}$}	&	\multicolumn{4}{|c}{SVM$_{TEXT}$}	&	\multicolumn{4}{|c}{BERT}		\\ \hline
Topic	&	\accuracy	&	\prec	&	\recall	&	\fscore	&	\accuracy	&	\prec	&	\recall	&	\fscore	&	\accuracy	&	\prec	&	\recall	&	\fscore	&	\accuracy	&	\prec	&	\recall	&	\fscore	\\ \hline
Kavanaugh	& 	83.5    &	83.2    &	83.2    & 	83.2    &	73.0	&	\textbf{83.3}	&	70.6	&	69.2	&	76.0	&	76.3	&	71.5	&	72.4	&	\textbf{84.7}	&	81.6	&	\textbf{83.8}	&	\textbf{82.7}	\\
Vaccine	& 	88.7    & 	73.1    &	93.8	& 	78.2    &	\textbf{88.2}	&	44.1	&	50.0	&	46.9	&	87.0	&	43.5	&	50.0	&	46.5	&	85.7	&	\textbf{98.0}	&	\textbf{85.7}	&	\textbf{91.4}	\\
Ilhan	&	 87.0   &	87.7    & 86.4     &	86.7    &	65.1	&	79.7	&	64.3	&	59.5	&	52.4	&	26.2	&	50.0	&	34.4	&	\textbf{87.9}	&	\textbf{91.1}	&	\textbf{86.4}	&	\textbf{88.7}	\\
Gun Control	& 92.3	&  91.8	&	 90.9  &	91.3    &	65.4	&	75.0	&	73.5	&	65.3	&	72.7	&	79.6	&	77.5	&	72.6	&	\textbf{93.8}	&	\textbf{97.6}	&	\textbf{93.0}	&	\textbf{95.2}	\\
Police Racism	& 	94.7    &   89.8	&	94.0	&	91.7	&	94.9	&	97.1	&	85.7	&	90.2	&	83.0	&	41.5	&	50.0	&	45.4	&	\textbf{96.0}	&	\textbf{98.3}	&	\textbf{96.7}	&	\textbf{97.5}	\\
Climate Change	&	95.8	&	93.8	&	95.3	&	94.5	&	82.9	&	90.9	&	62.5	&	65.0	&	81.6	&	90.2	&	62.5	&	64.6	&	\textbf{95.7}	&	\textbf{96.3}	&	\textbf{98.1}	&	\textbf{97.2}	\\
Midterm	&  85.9 &	83.5	&	84.1	&	83.8	&	87.3	&	87.4	&	82.7	&	84.4	&	69.8	&	75.6	&	77.9	&	69.7	&	\textbf{90.2}	&	\textbf{92.2}	&	\textbf{93.7}	&	\textbf{92.9}	\\
Immigration	&	84.4	&	83.8	&	84.0	&	83.	&	65.3	&	81.9	&	55.3	&	48.5	&	59.0	&	29.5	&	50.0	&	37.1	&	\textbf{89.5}	&	\textbf{91.2}	&	\textbf{91.2}	&	\textbf{91.2}	\\ \hline
Average 	& 89.0		&	85.8	&	88.9&	86.6	&	77.8	&	79.9	&	68.1	&	66.1	&	72.7	&	57.8	&	61.2	&	55.3	&	\textbf{90.4}	&	\textbf{93.3}	&	\textbf{91.1}	&	\textbf{92.1}	\\
    \end{tabular}
    \caption{Results on Sets A (no expansion). The best results in a row are in bold.}
    \label{tab:resSetA}
\end{table*}

\section{Experiments}
We split users in our test set on the basis of whether we were able to crawl their timelines or not.  Set A includes users for which we were not able to obtain their timeline tweets.  Set B includes users for which we were able to collect their timeline tweets.  We separated between them, because Set B would allow us to compare between setups that use timeline tweets with those that do not on identical users.

For Set A, we always trained on the training users with their on-topic tweets and the tested on the test users, who typically had less than 2 tweets on average.  We used four different classification setups, namely using fastText, SVM with retweeted accounts as features (SVM$_{RT}$), SVM with all words as features (SVM$_{TEXT}$), and fine-tuned BERT embeddings with a dense neural layer and softmax output (BERT).  We experimented with using the unsupervised method on Set A, but the unsupervised algorithm was not able to assign any test user to a cluster, mostly because the number of tweets and subsequently retweeted users per test user were too few.  For Set B, we experimented with the same classifiers using four different conditions, namely: not expanding either training or test sets with users' timeline tweets; expanding the test set only; expanding the training set only; and expanding both the training and test sets.

\section{Results and Discussion}
For all experiments, we report on per topic accuracy (\accuracy) and macro precision (\prec), recall (\recall), and F-measure (\fscore) across stances on a topic.  Table \ref{tab:resSetA} reports the results on Sets A where we were not able to expand the test set using timeline tweets.  As the results show, BERT yielded the best results in terms of \accuracy, \prec, \recall, and \fscore for most topics, with the highest overall averages across all scores. fastText trailed BERT, and SVM$_{TEXT}$ performed the worst.  This suggests that BERT, which uses contextual embeddings, is effective in performing accurate stance detection, even when classifying users with a very small number of topical tweets.  As for the Unsupervised method, using the unsupervised method was not able to assign any test user to a cluster, mostly because the number of tweets per test user were too few.  Hence, we omitted the unsupervised method from Table \ref{tab:resSetA}.

\begin{table}[ht]
    \centering
    \begin{tabular}{p{1.7cm}|c|p{4.5cm}}
        Error Type & No. & Examples \\ \hline
        Unexplained & 52 & \textit{Ilhan}: RIP CONS ... When they see me they take their hat off... I am DJ \newline \textit{Gun Control}: @BreitbartNews @NRA more lies \newline Climate: \$200 million a year could reverse climate change \\ \hline
        Vague & 48 &  \textit{Kavanaugh}: following the senate confirmation vote
 \newline \textit{Midterm}: RT\_@ali irresponsible for twitter \newline \textit{Police \& Racism}: after 40 yrs of reflection \\ \hline
        Quoting other side & 24 & \textit{Immigration}: this is a real crisis at the border \newline \textit{Kavanaugh}: why did jeff flake demand an investigation and then accept a bogus one
\\ \hline
        Sarcasm & 4 &  \textit{Immigration}: RT\_@infantry0300 someone should let ``Ms Hitler was a really great guy until he crossed the border into Poland''
\\
    \end{tabular}
    \caption{BERT error types with examples}
    \label{tab:BERTErrorAnalysis}
\end{table}

Table \ref{tab:resSetB} show the results on Set B, where we expanded the test, training, or either or both training and test user tweets using timeline tweets.  The results suggest the following:
\begin{itemize}[leftmargin=*]
    \item For BERT and fastText, which rely on the content of the tweets, we achieved the best results with no expansion or when we only expanded the training set.  The inclusion of non-topical tweets in the test set led to worse results overall.  We suspect that is happened because of the mismatch between training and test sets.  
    \item For SVM$_{RT}$ and Unsupervised classification, which rely exclusively on whom users retweeted, the expansion of the test dramatically improved overall \accuracy, \prec, \recall, and \fscore.  The positive improvement for both after timeline expansion suggests that the accounts that a user retweets are a strong signal of stance across multiple topics, and stances on multiple topics are likely correlated.  For example, a user who supported the Kavanaugh nomination was likely to vote republican in the midterm elections.  For future work, we plan to examine cross topic classification. 
    \item Similar to the results observed for Set A (\ref{tab:resSetA}), when no expansion is used, BERT led to the best overall results.  However, using unsupervised classification led to the best overall results across all setups, with expanding the test set only yielding slightly better results than expanding both the training and test sets. Expanding the test set only is significantly more efficient than expanding both training and test sets. 
    \item Using unsupervised classification failed to tag any of the users in the test set for any topic when the test set was not expanded, mostly because the number of tweets per test user and subsequent number of retweeted accounts were too few.
    \item SVM$_{TEXT}$ yielded the worse results overall, despite the inclusion of all the features in the tweets, such as retweeted accounts, hashtags, words, etc.  It seems that the inclusion of more features (compared to SVM$_{RT}$) confused the classifier leading to lower results.
    \item SVM$_{TEXT}$ and SVM$_{RT}$ led to the lowest results when we only expanded the training set.  For both setups, the classifier classified all users as belonging to one of the stances or the other.  Hence, \recall for one class was 100.0 and 0.0 for the other (with macro \recall = 50.0). We suspect that expanding the feature space in the training set confused the SVM classifier. Hence, both setups are not unusable. 
\end{itemize}

\begin{table*}[ht]
    \centering
    \scriptsize
    \begin{tabular}{l|c|c|c|c||c|c|c|c||c|c|c|c||c|c|c|c} 
	&	\multicolumn{16}{c}{fastText}	\\
	&	\multicolumn{4}{c}{No Expansion}	&	\multicolumn{4}{c}{Expanded Test only}  & \multicolumn{4}{c}{Expanded Test and Train} &	\multicolumn{4}{c}{Expanded Train only}	\\
Topic	&	\accuracy	 	&	P	&	R	&	F	&	\accuracy	&	\prec	&	\recall	&	\fscore	&	\accuracy	&	\prec	&	\recall	&	\fscore	&	\accuracy	&	\prec	&	\recall	&	\fscore	\\ \hline
Kavanaugh	&	\textbf{82.8}	&	\textbf{82.5} 	&	\textbf{82.3}	&	\textbf{82.4}	&	57.9	&	58.8	&	59.1	&	57.7	&	70.2	&	69.5	&	70.1	&	69.6	&	80.2	&	79.9	&	80.1	&	 80.0		\\
Vaccine	&	\textbf{86.4}	&	\textbf{68.3}	&	\textbf{85.0}	&	\textbf{72.3}	&	25.6	&	51.0	&	54.1	&	23.0	&	69.6	&	53.8	&	70.3	&	49.1	&	82.0	&	52.2	&	 52.7	&	 52.4	\\
Ilhan	&	 \textbf{85.9}	&	\textbf{89.4}	&	\textbf{85.3}	&	\textbf{85.4}	&	54.8	&	48.9	&	50.0	&	35.5	&	73.8	&	76.2	&	  72.0	&	71.9	&	84.4	&		84.3     	&	84.4	&	 84.3		\\
Gun Control	&	70.3	&	72.1	&	76.2	&	69.5	&	46.6	&	49.0	&	48.9	&	46.3	&	69.2	&	68.6	&	70.4	&	68.3	&	 \textbf{77.5}		&	\textbf{77.5}	&	 \textbf{78.6}		&	\textbf{76.0}	\\
Police Racism	&	\textbf{88.0}	&	  \textbf{81.8}	  	&	\textbf{90.3}	&	\textbf{84.5}	&	50.9	&	49.5	&	49.2	&	45.5	&	83.0	&	74.7	&	78.9	&	76.3	&	 \textbf{88.0}		&	 \textbf{81.0}		&	\textbf{83.1}	&	 \textbf{82.0}		\\
Climate Change	&	 75.6		&	 68.1		&	79.6	&	68.0	&	78.2	&	51.3	&	51.0	&	50.9	&	80.3	&	68.6	&	80.3	&	71.0	&	\textbf{87.6}	&	\textbf{80.7}	&	 \textbf{87.6}		&	 \textbf{83.2}		\\
Midterm	&	84.9	&	79.4	&	82.2	&	80.6	&	42.6	&	49.5	&	49.3	&	41.7	&	81.0	&	 74.5	&	77.4	&	75.7	&	\textbf{91.7}	&	\textbf{91.0}	&	\textbf{88.5}	&	\textbf{89.6}	\\
Immigration	&	\textbf{87.5}	&	\textbf{87.7}	&	\textbf{87.7}	&	\textbf{87.5}	&	49.7	&	50.1	&	50.1	&	 49.5	&	79.0	&	78.4	&	79.0	&	 78.6		&	 83.0		&	84.3	&	84.3	&	 83.0	\\ \hline
Average 	&	82.6	&	\textbf{78.6}	&	\textbf{83.5}	&	\textbf{78.9}	&	50.8	&	51.0	&	51.5	&	43.8	&	65.8	&	70.5	&	74.8	&	70.1	&	\textbf{84.3}	&	 \textbf{78.6}		&	 79.9		&	\textbf{78.8}	\\
Std Dev	&	6.0	&	7.8	&	4.2	&	7.0	&	13.9	&	3.1	&	3.3	&	9.9	&	5.3	&	7.2	&	4.2	&	8.6	&	4.3	&	10.8	&	10.8	&	10.6	\\

		\hline\hline															
	&	\multicolumn{16}{c}{SVM$_{RT}$} \\
&	\multicolumn{4}{c}{No Expansion}	&	\multicolumn{4}{c}{Expanded Test only}  & \multicolumn{4}{c}{Expanded Test and Train} &	\multicolumn{4}{c}{Expanded Train only}	\\
Topic	&	\accuracy	&	\prec	&	\recall	&	\fscore	&	\accuracy	&	\prec	&	\recall	&	\fscore	&	\accuracy	&	\prec	&	\recall	&	\fscore	&	\accuracy	&	\prec	&	\recall	&	\fscore	\\ \hline
Kavanaugh	&	73.7	&	78.9	&	72.5	&	71.7	&	84.9	&	84.9	&	83.8	&	84.2	&	\textbf{86.8}	&	\textbf{88.7}	&	\textbf{84.8}	&	\textbf{85.8}	&	52.6	&	26.3	&	50.0	&	34.5	\\
Vaccine	&	95.5	&	47.7	&	50.0	&	48.8	&	\underline{\textbf{96.4}}	&	\textbf{98.2}	&	\textbf{75.0}	&	\textbf{82.4}	&	\textbf{96.4}	&	\textbf{98.2}	&	\textbf{75.0}	&	\textbf{82.4}	&	95.5	&	47.7	&	50.0	&	48.8	\\
Ilhan	&	57.1	&	77.3	&	55.9	&	45.8	&	\textbf{89.5}	&	\textbf{91.7}	&	\textbf{88.9}	&	\textbf{89.2}	&	71.1	&	82.3	&	69.4	&	67.2	&	51.4	&	25.7	&	50.0	&	34.0	\\
Gun Control	&	66.7	&	72.7	&	76.9	&	66.3	&	75.0	&	\textbf{78.6}	&	\textbf{81.3}	&	74.8	&	\textbf{79.2}	&	76.9	&	75.0	&	\textbf{75.8}	&	72.2	&	36.1	&	50.0	&	41.9	\\
Police Racism	&	90.6	&	94.6	&	78.6	&	83.5	&	\textbf{92.3}	&	88.0	&	\textbf{92.1}	&	\textbf{89.7}	&	90.4	&	\textbf{94.4}	&	79.2	&	83.9	&	78.1	&	39.1	&	50.0	&	43.9	\\
Climate Change	&	92.3	&	96.0	&	70.0	&	76.5	&	\underline{\textbf{100.0}}	&	\underline{\textbf{100.0}}	&	\underline{\textbf{100.0}}	&	\underline{\textbf{100.0}}	&	96.2	&	97.9	&	85.7	&	90.6	&	87.2	&	43.6	&	50.0	&	46.6	\\
Midterm	&	\underline{\textbf{95.5}}	&	\underline{\textbf{97.2}}	&	90.0	&	\textbf{93.0}	&	92.9	&	89.3	&	\textbf{91.9}	&	90.5	&	88.1	&	84.3	&	81.9	&	83.0	&	77.3	&	38.6	&	50.0	&	43.6	\\
Immigration	&	59.5	&	78.6	&	55.9	&	46.9	&	\textbf{97.6}	&	\textbf{97.5}	&	\underline{\textbf{97.8}}	&	\textbf{97.6}	&	73.8	&	79.4	&	71.5	&	71.0	&	54.1	&	27.0	&	50.0	&	35.1	\\ \hline
Average 	&	78.9	&	80.4	&	68.7	&	66.6	&	\textbf{91.1}	&	\textbf{91.0}	&	\textbf{88.8}	&	\textbf{88.6}	&	85.2	&	87.8	&	77.8	&	80.0	&	71.0	&	35.5	&	50.0	&	41.0	\\
Std Dev	&	15.4	&	15.3	&	12.8	&	16.7	&	7.5	&	6.9	&	7.9	&	7.6	&	9.0	&	7.8	&	5.7	&	7.4	&	15.7	&	7.8	&	0.0	&	5.4	\\

		\hline\hline															
	&	\multicolumn{16}{c}{SVM$_{TEXT}$} \\
&	\multicolumn{4}{c}{No Expansion}	&	\multicolumn{4}{c}{Expanded Test only}  & \multicolumn{4}{c}{Expanded Test and Train} &	\multicolumn{4}{c}{Expanded Train only}	\\
Topic	&	\accuracy	&	\prec	&	\recall	&	\fscore	&	\accuracy	&	\prec	&	\recall	&	\fscore	&	\accuracy	&	\prec	&	\recall	&	\fscore	&	\accuracy	&	\prec	&	\recall	&	\fscore	\\ \hline
Kavanaugh	&	68.5	&	67.8	&	67.0	&	67.2	&	\textbf{70.9}	&	\textbf{71.3}	&	\textbf{69.0}	&	\textbf{69.2}	&	56.4	&	65.9	&	60.4	&	53.8	&	42.6	&	21.3	&	50.0	&	29.9	\\
Vaccine	&	\textbf{92.9}	&	\textbf{46.4}	&	50.0	&	\textbf{48.2}	&	89.5	&	46.4	&	48.1	&	47.2	&	49.1	&	52.9	&	\textbf{61.1}	&	40.2	&	7.1	&	3.6	&	50.0	&	6.7	\\
Ilhan	&	52.6	&	26.3	&	50.0	&	34.5	&	53.9	&	26.9	&	50.0	&	35.0	&	\textbf{71.8}	&	\textbf{72.2}	&	\textbf{72.2}	&	\textbf{71.8}	&	47.4	&	23.7	&	50.0	&	32.1	\\
Gun Control	&	56.0	&	72.5	&	65.6	&	54.8	&	\textbf{69.2}	&	\textbf{76.5}	&	\textbf{76.5}	&	\textbf{69.2}	&	46.2	&	59.1	&	56.2	&	44.9	&	36.0	&	18.0	&	50.0	&	26.5	\\
Police Racism	&	77.4	&	38.7	&	50.0	&	43.6	&	\textbf{85.2}	&	\textbf{81.0}	&	\textbf{72.6}	&	\textbf{75.5}	&	66.7	&	61.9	&	66.7	&	61.4	&	22.6	&	11.3	&	50.0	&	18.5	\\
Climate Change	&	\textbf{92.5}	&	\textbf{96.0}	&	\textbf{71.4}	&	\textbf{77.9}	&	79.6	&	43.0	&	45.7	&	44.3	&	42.6	&	59.2	&	67.0	&	41.0	&	13.2	&	6.6	&	50.0	&	11.7	\\
Midterm	&	61.4	&	68.5	&	75.0	&	60.4	&	80.0	&	71.3	&	62.1	&	64.0	&	\textbf{84.4}	&	\textbf{78.3}	&	\textbf{86.4}	&	\textbf{80.6}	&	22.7	&	11.4	&	50.0	&	18.5	\\
Immigration	&	54.8	&	27.4	&	50.0	&	35.4	&	58.1	&	62.1	&	53.2	&	45.0	&	\textbf{86.1}	&	\textbf{87.7}	&	\textbf{84.8}	&	\textbf{85.4}	&	54.8	&	27.4	&	50.0	&	35.4	\\ \hline
Average 	&	69.5	&	55.5	&	59.9	&	52.7	&	\textbf{73.3}	&	\textbf{59.8}	&	\textbf{59.7}	&	\textbf{56.2}	&	62.9	&	67.2	&	69.3	&	59.9	&	30.8	&	15.4	&	50.0	&	22.4	\\
Std Dev	&	15.4	&	23.1	&	10.2	&	14.3	&	11.8	&	17.8	&	11.2	&	14.0	&	15.9	&	10.8	&	10.4	&	16.7	&	15.9	&	7.9	&	0.0	&	9.5	\\
		\hline\hline															
	&	\multicolumn{16}{c}{BERT} \\
&	\multicolumn{4}{c}{No Expansion}	&	\multicolumn{4}{c}{Expanded Test only}  & \multicolumn{4}{c}{Expanded Test and Train} &	\multicolumn{4}{c}{Expanded Train only}	\\
Topic	&	\accuracy	&	\prec	&	\recall	&	\fscore	 	&	\accuracy	&	\prec	&	\recall	&	\fscore	&	\accuracy	&	\prec	&	\recall	&	\fscore	&	\accuracy	&	\prec	&	\recall	&	\fscore	\\ \hline
Kavanaugh	&	\textbf{82.5}	&	\textbf{79.}	&	\textbf{81.8}	&	\textbf{80.7}	&	60.9	&	51.1	&	69.5	&	58.9	&	70.9	&	62.2	&	71.0	&	66.3	&	80.5	&	76.9	&	78.9	&	77.9	\\
Vaccine	&	89.1	&	\textbf{99.0}	&	\textbf{88.7}	&	93.6	&	38.9	&	96.7	&	37.5	&	54.0	&	73.5	&	98.3	&	73.6	&	84.2	&	\textbf{90.9}	&	98.2	&	\textbf{91.7}	&	\textbf{94.8}\\
Ilhan	&	91.6	&	\textbf{93.6}	&	\textbf{90.7}	&	92.1	&	61.7	&	62.7	&	74.4	&	68.1	&	76.0	&	72.9	&	89.5	&	80.4	&	\textbf{93.0}	&	92.1	&	\textbf{94.6}	&	\textbf{93.3}	\\
Gun Control	&	\textbf{86.2}	&	88.9	&	\textbf{95.0}	&	\textbf{83.5}	&	88.8	&	70.8	&	60.1	&	65.0	&	70.4	&	83.9	&	67.3	&	74.7	&	73.0	&	\textbf{90.0}	&	69.2	&	78.3	\\
Police Racism	&	\textbf{93.0}	&	\textbf{96.7}	&	94.4	&	\textbf{95.5}	&	67.6	&	86.0	&	70.7	&	77.6	&	83.4	&	93.6	&	84.9	&	89.0	&	91.6	&	93.9	&	\textbf{95.4}	&	94.7	\\
Climate Change	&	\textbf{94.1}	&	\textbf{97.4}	&	\textbf{95.0}	&	\textbf{96.2}	&	89.0	&	75.0	&	75.0	&	81.4	&	71.0	&	92.8	&	71.6	&	80.8	&	80.5	&	93.3	&	82.4	&	87.5	\\
Midterm	&	91.7	&	93.3	&	95.1	&	94.2	&	57.8	&	81.6	&	57.9	&	67.7	&	82.3	&	92.2	&	84.0	&	87.9	&	\textbf{94.3}	&	\textbf{95.1}	&	\textbf{97.5}	&	\textbf{96.3}	\\
Immigration	&	88.1	&	92.4	&	86.0	&	89.1	 	&	65.3	&	72.9	&	65.3	&	68.9	&	81.3	&	84.3	&	83.8	&	84.0	&	\textbf{92.6}	&	\textbf{97.4}	&	\textbf{88.4}	&	\textbf{92.7}	\\ \hline
Average 	&	\textbf{89.6}	&	\textbf{93.4}	&	\textbf{89.4}	&	\textbf{91.3}	&	62.4	&	74.6	&	63.8	&	67.7	&	76.1	&	85.0	&	78.2	&	80.9	&	87.0	&	92.1	&	87.3	&	89.4	\\
Std Dev	&	8.8	&	7.2	&	11.4	&	8.6	&	12.9	&	13.1	&	11.5	&	8.4	&	5.1	&	11.4	&	7.7	&	7.0	&	7.4	&	6.3	&	9.1	&	7.0	\\
	\hline\hline			
	&	\multicolumn{16}{c}{Unsupervised} \\
&	\multicolumn{4}{c}{No Expansion}	&	\multicolumn{4}{c}{Expanded Test only}  & \multicolumn{4}{c}{Expanded Test and Train} &	\multicolumn{4}{c}{Expanded Train only}	\\
Topic	&	\multicolumn{4}{c||}{}	&	\accuracy	&	\prec	&	\recall	&	\fscore	 	&	\accuracy	&	\prec	&	\recall	&	\fscore	 	&	\multicolumn{4}{c}{}	\\ \hline
Kavanaugh	& \multicolumn{4}{c||}{}	&	84.6	&	84.2	&	84.2	&	84.2	&	\underline{\textbf{90.4}}	&	\underline{\textbf{89.2}}	&	\underline{\textbf{91.3}}	&	\underline{\textbf{89.9}}	& \multicolumn{4}{c}{}	\\ 
Vaccine	& \multicolumn{4}{c||}{}	&	\textbf{96.3}	&	83.3	&	\underline{\textbf{99.0}}	&	\textbf{89.5}	&	95.3	&	\textbf{97.1}	&	75.0	&	81.8	& \multicolumn{4}{c}{}	\\
Ilhan	& \multicolumn{4}{c||}{}	&	\textbf{91.9}	&	\textbf{91.6}	&	\textbf{92.1}	&	\textbf{91.8}	&	\textbf{91.9}	&	\textbf{91.6}	&	\textbf{92.1}	&	\textbf{91.8}	& \multicolumn{4}{c}{}	\\
Gun Control	& \multicolumn{4}{c||}{}	&	\underline{\textbf{95.8}}	&	\underline{\textbf{90.6}}	&	\underline{\textbf{86.4}}	&	\underline{\textbf{86.9}}	&	\underline{\textbf{87.5}}	&	\underline{\textbf{90.6}}	&	\underline{\textbf{86.4}}	&	\underline{\textbf{86.9}}	& \multicolumn{4}{c}{}	\\
Police Racism	& \multicolumn{4}{c||}{}	&	\underline{\textbf{100.0}}	&	\underline{\textbf{100.0}}	&	\underline{\textbf{100.0}}	&	\underline{\textbf{100.0}}	&	98.0	&	98.7	&	95.8	&	97.2	& \multicolumn{4}{c}{}	\\
Climate Change	& \multicolumn{4}{c||}{}	&	\textbf{96.2}	&	\textbf{97.8}	&	\textbf{88.9}	&	\textbf{92.6}	&	\textbf{92.3}	&	\textbf{96.7}	&	\textbf{85.0}	&	\textbf{89.5}	& \multicolumn{4}{c}{}	\\
Midterm	& \multicolumn{4}{c||}{}	&	\underline{\textbf{100.0}}	&	\underline{\textbf{100.0}}	&	\underline{\textbf{100.0}}	&	\underline{\textbf{100.0}}	&	\textbf{95.0}	&	\textbf{96.8}	&	\textbf{90.9}	&	\textbf{93.3}	& \multicolumn{4}{c}{}	\\
Immigration	& \multicolumn{4}{c||}{}	&	\underline{\textbf{100.0}}	&	\underline{\textbf{100.0}}	&	\underline{\textbf{100.0}}	&	\underline{\textbf{100.0}}	 	&	\underline{\textbf{97.6}}	&	\underline{\textbf{97.8}}	&	\textbf{97.5}	&	\underline{\textbf{97.6}}	 	& \multicolumn{4}{c}{}	\\ \hline
Average 	& \multicolumn{4}{c||}{}	&	\underline{\textbf{95.6}}	&	93.4	&	\underline{\textbf{93.8}}	&	\underline{\textbf{93.1}}	&	93.5	&	\underline{\textbf{94.8}}	&	89.3	&	91.0	& \multicolumn{4}{c}{}	\\
Std Dev 	& \multicolumn{4}{c||}{}	&	4.9	&	6.6	&	6.3	&	5.9	&	3.4	&	3.5	&	6.7	&	4.9	& \multicolumn{4}{c}{}	\\
    \end{tabular}
    \caption{Results on Set B with and without expansion of either training or test sets. Highest \accuracy, \prec, \recall, and \fscore per method are bolded, and highest values overall are underlined. The table reports the average and standard deviation of per topic scores.}
    \label{tab:resSetB}
\end{table*}

We computed the standard deviation (SD) of all our measures across topics for every setup. Lower SD coupled with high \accuracy and \fscore is desirable as they indicate the setup produces consistently high results across topics.  Unsupervised classification yielded the lowest SD values and highest overall score.  BERT and fastText with no expansion and SVM$_{RT}$ with expanded test set had slightly higher SD.

Thus, if we are able to scrape a user's timeline tweets, it is advantageous to use a method that relies on which accounts the user retweeted, with unsupervised classification producing the best results.  As we will show in the error analysis, the success of unsupervised classification is contingent on users retweeting a sufficient number of times, particularly politically related accounts in our case.  When timeline tweets are not available, it is best to use contextualized embeddings, such as BERT, to represent tweets and subsequently classify them.

\subsection{Error Analysis}

We analyzed all the errors in Set B that were produced by BERT with no expansion, as it represents the best results when expansion is not possible, and those produced by unsupervised classification with the expansion of the test set only, as this produced the best overall results. Since we used BERT to perform tweet-level classification, we manually inspected all 129 misclassified tweets across all topics.  Generally we found four types of errors, namely: unexplainable errors where the tweets clearly expressed stance, but the classifier mislabeled them; vague tweets that have no clear clues; tweets in which the user uses the language of the opposing side; and sarcastic tweets. Table \ref{tab:BERTErrorAnalysis} lists the error types with their frequencies and provides example tweets.

For unsupervised classification, we manually examined all 15 users that were missclassified, of whom 7 were from the Kavanaugh topic.  The prominent reasons for incorrect classification were: 
\begin{itemize}
    \item Lack of sufficient retweets for a user.  The percentage of retweets ranged between 1-1.4\% of all tweets for three of the misclassified users (2 for \textit{climate change} and 1 for \textit{Kavanaugh}).  Yet for another user (\textit{Kavanaugh}), their timeline tweets were only 137 with 28 retweets.
    \item Geographic mislabeling, where 2 accounts were not US accounts (1 for \textit{Ilhan} and 1 for \textit{Kavanaugh}).
    \item users retweeting mostly apolitical accounts such as music, art, or cars related accounts (1 for \textit{climate change}, 1 for \textit{vaccine}, and 3 for \textit{Kavanaugh}). Retweeting of politically biased accounts and media sources seem to provide strong signals for classification.
    \item User goes against the general opinion of his group.  Specifically, there was a clearly republican user who was criticizing the National Rifle Association (NRA) (\textit{gun control}).
\end{itemize}

Thus, the most common reason for misclassification was the dearth of retweets from politically oriented or topically related accounts.



\section{Conclusion}
In this paper, we presented two methods for classifying users according to their stance towards a target.  The first utilizes contextualized embeddings to represent tweets and then uses deep neural network for classification.  This approach led to results that outperforming two strong baselines.  The second utilizes additional tweets from users' timelines to cluster test users with other users with known labels in an unsupervised manner.  The first method yielded the best results when timeline tweets were not available, while the second yielded even better results overall.  Given the overall setup described in the paper, where the training data was obtained using unsupervised user classification, we can automatically label the most active users with nearly perfect accuracy, and we can label users with only few topical tweets with high accuracy, often above 95\% when we can obtain their timeline tweets.  For future work, we plan to explore the effectiveness of cross topic classification, where training and testing are done on different topics.  Perhaps, we can build unified models that could be used across multiple topics for a given a population of Twitter users (ex. users who are interested in US politics).

\bibliographystyle{aaai}
\bibliography{references}

\end{document}